\documentclass{emulateapj}

\shorttitle{Gas flows in mergers}
\shortauthors{Torres-Flores et al.}

\begin{document}


\title{Witnessing gas mixing in the metal distribution of the \\Hickson Compact Group HCG 31}


\author{S. Torres-Flores}
\affil{Departamento de F\'isica, Universidad de La Serena, Av. Cisternas 1200, La Serena, Chile}
\email{storres@dfuls.cl}

\author{C. Mendes de Oliveira}
\affil{Instituto de Astronomia, Geof\'isica e Ci\^encias Atmosf\'ericas da Universidade de S\~{a}o Paulo, Cidade Universit\'aria, CEP:05508-900, S\~{a}o Paulo, SP, Brazil}

\author{P. Amram}
\affil{Aix Marseille Universit\'e, CNRS, LAM (Laboratoire d'Astrophysique de Marseille) UMR 7326, 13388, Marseille, France}

\author{M. Alfaro-Cuello}
\affil{Departamento de F\'isica, Universidad de La Serena, Av. Cisternas 1200, La Serena, Chile}

\author{E. R. Carrasco}
\affil{Gemini Observatory/AURA, Southern Operations Center, Casilla 603, La Serena, Chile}

\author{D. F. de Mello}
\affil{Catholic University of America, Washington, DC 20064, USA}


\altaffiltext{}{Based on observations obtained at the Gemini Observatory, which is operated by the Association of Universities for Research in Astronomy, Inc., under a cooperative agreement with the NSF on behalf of the Gemini partnership: the National Science Foundation (United States), the Science and Technology Facilities Council (United Kingdom), the National Research Council (Canada), CONICYT (Chile), the Australian Research Council (Australia), Minist\'erio da Ci\^encia e Tecnologia (Brazil) and Ministerio de Ciencia, Tecnolog\'ia e Innovaci\'on Productiva (Argentina) -- Observing run: GS-2012B-Q-60.}


\begin{abstract}
We present, for the first time, direct evidence that, in a merger of disk galaxies, the preexisting central metallicities will mix as a result of gas being transported in the merger interface region, along the line that joins the two coalescing nuclei. This is shown using detailed 2D kinematics as well as metallicity measurements for the nearby ongoing merger in the center of the compact group HCG 31. We focus on the emission line gas, which is extensive in the system. The two coalescing cores display similar oxygen abundances while in between the two nuclei the metallicity changes smoothly from one nucleus to the other, indicating a mix of metals in this region, which is confirmed by the high resolution H$\alpha$ kinematics (R=45900). This nearby system is especially important because it involves the merging of two fairly low mass and clumpy galaxies (LMC-like galaxies), making it an important system for comparison with high redshift galaxies.
\end{abstract}





\section{Introduction}

The most accepted scenario for the evolution of systems of merging galaxies predicts that large-scale gas flows are widespread and they may occur already at first passage (e.g. Mihos \& Hernquist 1996, Rupke et al. 2010a). Theoretical studies predict that, in major mergers, tidal torques will develop bars and will induce gas to lose angular momentum and flow towards the center, fueling massive central starbursts, active galactic nuclei and/or quasar activity (e.g. Barnes and Hernquist 1996).  As shown by Torrey et al.  (2012) and references therein, these inflows could either cause a depression in the nuclear metallicity of gas-poor disk-disk interactions or cause an enhancement in the central metallicity due to star formation in gas-rich disk-disk interactions, given that the metallicity of the  merger remnant is mainly set by the competition between the inflow of low-metallicity gas and enrichment from star formation. However, other factors such as gas consumption and galactic winds also contribute as secondary players (Torrey et al. 2012).

Although these large inflows in merging galaxies have been expected from simulations and theory, they have never been observationally witnessed mainly due to the lack of detailed kinematic data over a large extent of colliding systems and with sufficiently high spatial and spectral resolutions.  In particular, in order to identify the various intertwined velocity components of the inflowing gas, one needs 2D kinematic data over an extended area that is usually not delivered by IFUs and with sufficiently high spectral resolution of typically \mbox{R $>$ 30000}. But even if one had in hands high resolution wide-field 2D velocity maps for a sample of mergers and close pairs, it would be still generally quite difficult to disentangle the individual gas kinematic components because signatures of AGN activity, shocked regions and winds are often present, particularly in the case of strong inflows (Rich et al. 2011). AGN activity and shocks can blur the signatures of the several components of the gas kinematics by broadening the lines, erasing gas peaks and consequently mixing or altering the information.

According to the simulations of Mihos and Hernquist (1996), gaseous inflows are strongest when the colliding galaxies have dense central bulges and they are in the final stages of merging, but these are strongly affected specially by shocks and AGN activity, which may cover up the kinematic inflow signatures and may difficult interpretation of the data. Inflows in bulgeless galaxies, however, although they are expected to be weaker, they tend to occur earlier in the interaction and they are much less affected by shocks. Our best bet for directly witnessing gas flows in a colliding system is then to focus on the gas kinematic signatures of mergers of low-mass, late-type, disk-disk progenitors, where bulges will be less important and AGN activity, shocks and winds may play less of a role.

Here we study an ideal group in which gas flows may be detected: the system HCG 31A and HCG 31C in the compact group HCG 31, which are a low-mass (both galaxies have masses similar to that of the Large Magellanic Cloud), low-metallicity (where member HCG 31C has a value of 12+log(O/H) = 8.22, L\'opez-S\'anchez et al. 2004) and low-separation (projected separation of 3.5 arcsec or 0.9 kpc between the two merging nuclei). Amram et al. (2007) have found that the two merging nuclei are in a bound orbit, with almost parallel spin axes, like a set of gear wheels, in a prograde encounter. The two merging cores have had at least one earlier passage and they have a high star-formation rate (10.6 M$_{\odot}$ yr$^{-1}$, Gallagher et al. 2010), being in an early stage of merger.

The main goal of this Letter is to show how the kinematics of the HCG 31AC system relates to its metallicity in the merger interface region, along the line that joins the two merging nuclei. At the redshift of HCG 31 (z=0.013473, NED database), 1''=0.26 kpc (for H$_{0}$=75 km s$^{-1}$ Mpc$^{-1}$). The Letter is organized as follows: in \S \ref{observations} we describe the observations and data reduction. In \S \ref{results} we present the results. Finally, a discussion and a summary of our results are presented in \S \ref{summary}.
    
\section{Observations and data reduction}
\label{observations}

\subsection{Gemini/GMOS IFU data}

The central region of HCG 31 has been observed with the Integral Field Unit (IFU, Allington-Smith et al. 2002) of the Gemini MultiObject Spectrograph (GMOS, Hook et al. 2004), at the Gemini South Observatory, under the programme GS-2012B-Q-60. Three fields of view (see Figure \ref{pointing}) were observed by using the R400 grating and the {\textit{r'}-band filter and for each field we observed three exposures of 700 sec, with a mean seeing of 0.4''. The grating was centered at 6300 {\AA}, using the two slit mode, covering a spatial area of 5''$\times$7'' for each field, from 5620 {\AA} to 6960 {\AA}. Data reduction was performed by using the Gemini package, in {\sc iraf}. The 2D images were transformed to 3D data cube and re-sampled as square pixels with 0.1" spatial resolution and corrected by differential atmospheric refraction. The data was flux calibrated by using the standard star LTT 4364. 

Each spectrum of the data cube has been corrected by Galactic extinction, using E(B-V)=0.04 mag (NED database) and the Fitzpatrick (1999) extinction law. Internal extinction was corrected by using the Calzetti et al. (2000) extinction and an average nebular colour excess of E(B-V)=0.08 mag, which was estimated for the central region of HCG 31AC by using the spectra published by Mendes de Oliveira et al. (2006). This approach was used because our IFU data do not include the H$\beta$ emission line. Emission line fluxes were estimated with the interactive routine {\sc fluxer} in {\sc idl} (code written by Christof Iserlohe). 

\begin{figure}
\includegraphics[width=0.5\textwidth]{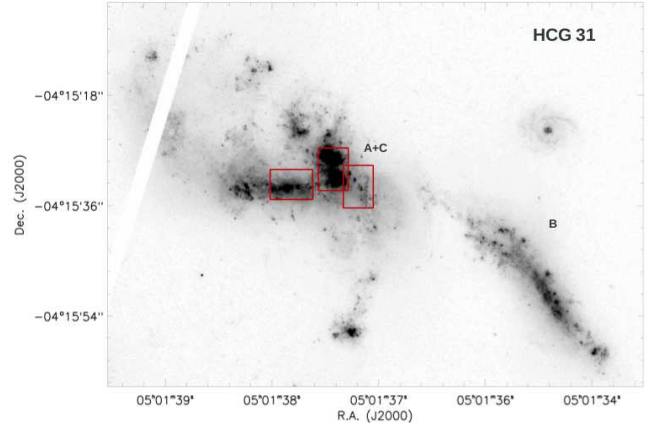}
\caption{\textit{Hubble Space Telescope} (\textit{HST}) optical image (ACS, F435W) of the compact group HCG 31. The image shows the regions A+C, B and part of the southeast tidal tail. The red squares show the three different regions where the IFU are located.}
\label{pointing}
\end{figure}

\subsection{Fabry-Perot data}

Observations were carried out in August 2000, at the ESO 3.6 m telescope, in Chile, using the Fabry-Perot instrument CIGALE. A total of 48 channels were mapped, with a free spectral range of 155 km s$^{-1}$, which produced a sample step of 3.2 km s$^{-1}$. The spectral resolution of this data set is R=45900 (see Amram et al. 2007). The pixel size is 0.405 arcsec pixel$^{-1}$ and the observations were taken with a mean seeing of 0.7 arcsec.

\section{Results}
\label{results}  

\begin{figure*}
\includegraphics[width=1.0\textwidth]{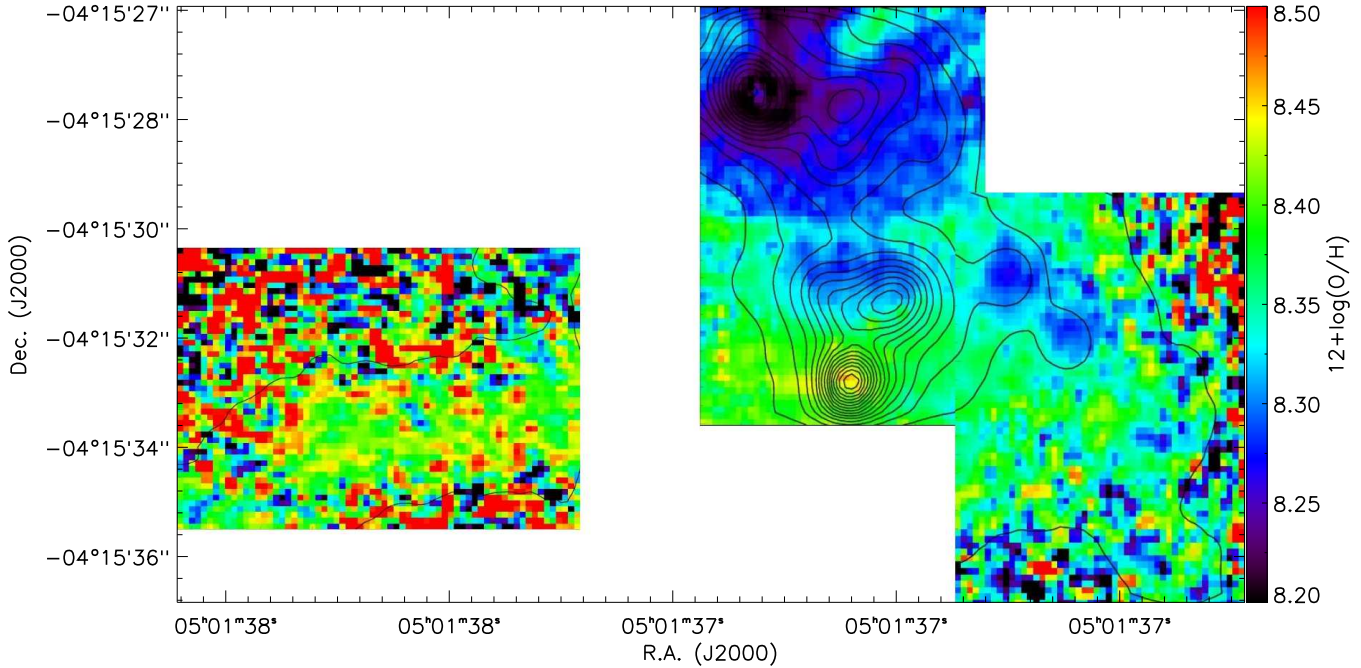} 
\caption{Oxygen abundance map for the central region of HCG 31. This has been derived from the N2 calibrator following Marino et al. (2013). Black contours represent the H$\alpha$ emission in HCG 31.}
\label{abundances}
\end{figure*}

\subsection{Determining the oxygen abundances}

Given the spectral coverage of our observations, oxygen abundances were estimated by using the strong line abundance indicator N2 which is defined as \mbox{N2= log([NII]$\lambda$6584/H$\alpha$$\lambda$6563)}. We used the calibrator recently revised by Marino et al. (2013). We do not correct the spectra by Balmer absorption that may have been related with an underlying old population given the very weak continuum (L\'opez-S\'anchez et al. 2004).  We note that the empirical calibrations are based on oxygen abundances derived from the direct method and they are affected by uncertainties related with the scatter on the calibration itself.  Also the use of different calibration methods on the same observational data set can produce differences in the estimated values. However, as pointed out by Bresolin et al. (2012), differential analysis are useful to minimize these discrepancies. In this analysis, uncertainties in the oxygen abundances were first estimated by propagating the flux uncertainties (derived from the task {\sc fluxer}) and considering the standard deviation in the zero-point and slope of the calibrator regression. These together amount to tipically 0.03 mag. However, by far the largest source of error is the 0.16 dex scatter in the N2 calibration found by Marino et al.  (2013). All these uncertainties are added in quadrature. We also attempted to derive the metallicity gradient using the N2S2 prescription of P\'erez-Montero (2014), but given the much lower fluxes of the used lines, the error bars are large. A third determination of the metallicity was employed using the GMOS data of HCG 31 published by Mendes de Oliveira et al. (2006). These authors obtained a spectrum  through a slit placed across members HCG 31AC. These have been corrected by Galactic and internal extinction by using the same extinction laws used for the IFU data, and using the observed H$\alpha$/H$\beta$ ratio for each extraction, with the corresponding intrinsic ratio taken from Osterbrock (1989), for T$_{e}$ = 10 000 K and N$_{e}$ = 100. The main nebular emission lines present in this latter data set are H$\beta$, [OIII] 4959 {\AA}, [OIII] 5007 {\AA}, [NII] 6548 {\AA}, H$\alpha$, [NII] 6584 {\AA}, and therefore, oxygen abundances were estimated with the O3N2 indicator suggested by Marino et al. (2013).

In Fig. \ref{abundances} we show the oxygen abundance map of the central regions of HCG 31 (derived from the N2 calibrator), corresponding to the areas of the three red boxes of Fig. \ref{pointing}. We find that the star-forming complex associated with HCG 31C has an oxygen abundance of \mbox{12+log(O/H)=8.22$\pm$0.17}, while HCG 31A has an abundance of \mbox{12+log(O/H)=8.44$\pm$0.16}. We note that the latter value was also measured along the bar-like structure that forms the eastern part of HCG 31A. The absolute numbers here are not important, given that they may change with calibrator. 

Inspecting the central field of Figure \ref{abundances} it can be noted that there is a smooth gradient between the two main starbursting complexes. Taking into account the uncertainties, the metallicities of these two complexes are similar, which suggests that the strong gravitational encounter between HCG 31A and HCG 31C is currently mixing their chemical content. It is in agreement with the general notion that strong interactions flatten metallicity gradients.

In order to emphasize this result, in Figure \ref{gradient} we have plotted with filled circles the oxygen abundance gradient along the region that connects the strong star forming blobs of HCG 31C to those of HCG 31A, covering a very similar area (but smaller) to that shown by the seven green boxes in Fig. \ref{profiles} (see description of this figure in \S 3.2). This gradient was derived from the 2D abundance maps using the N2 method (in this case we mimic a long slit observation). We also show in Fig. \ref{gradient} metallicity gradients using the other calibrators, O3N2 and N2S2 (filled triangles and empty stars, respectively). We see a smooth transition in the oxygen abundance gradient as measured by the N2 calibrator. The N2S2 calibrator shows the same trend, however, the scatter in this calibrator is large given the smaller S/N ratios of the [SII] lines (and for this reason we have binned these data in Fig. \ref{gradient}). In the case of O3N2, the S/N ratio are also poorer given that these come from a long-slit observation, with few data points compared to the IFU determination of the N2 calibrator. Considering the uncertainties in the oxygen abundances, we find that the metallicities of HCG 31A and HCG 31C are compatible and the merger interface region displays a smooth gradient in the chemical abundance. Using a linear fit on the N2 measurements, we derived a gradient of 1.5$\times$10$^{-4}$ dex pc$^{-1}$ for the oxygen abundances in HCG 31AC, which reflects that the metallicities of both systems are compatible. Given that we do not detect any discontinuity in the gradient, we can speculate that the smooth gradient is real. This indicates that gas mixing is taking place between these galaxies. This result is confirmed by kinematics as we show in the following section.

\begin{figure}
\includegraphics[width=0.47\textwidth]{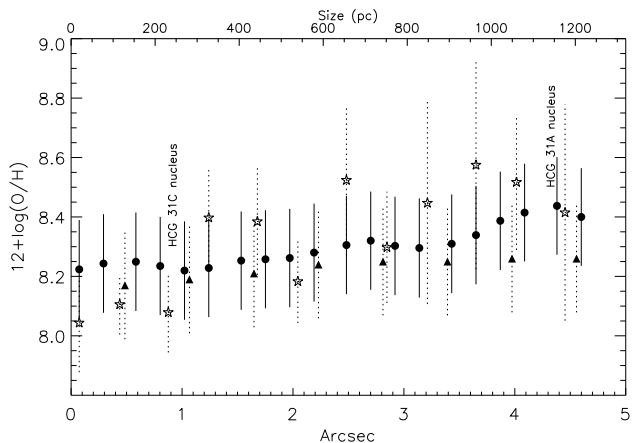} 
\caption{Oxygen abundance gradient extracted from the map shown in Fig. \ref{abundances}. This gradient was derived by simulating a long slit passing through the seven green boxes displayed in Fig. \ref{profiles}. Black filled circles, black filled triangles and empty stars represent the oxygen abundances derived from the N2, O3N2 and N2S2 methods, respectively. This gradient was estimated along the two main star forming nuclei of HCG 31C and HCG 31A, respectively. In the case of the N2 gradient the sampling was chosen to match the seeing and in the case of the N2S2 calibrator we have binned the data given the low S/N ratio of the [SII] lines.}
\label{gradient}
\end{figure}

\begin{figure*}
\includegraphics[width=\textwidth]{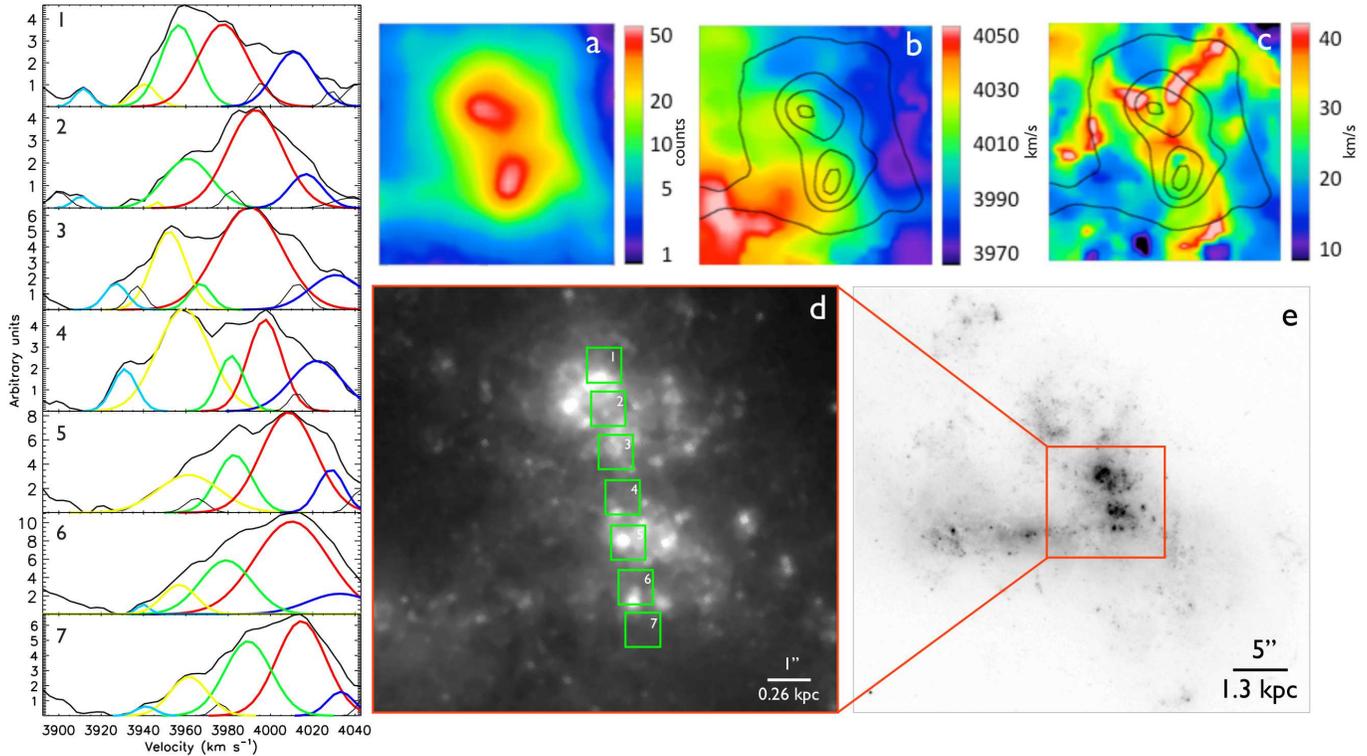}
\caption{(a) Total flux (in arbitrary units) within the interference filter of the Fabry-Perot, around H$\alpha$ (20 Angstrons), roughly matching the \textit{HST} image displayed in (d), shown in logarithm scale. (b) and (c), adapted from Amram et al.  (2007), are the velocity field and velocity dispersion maps corresponding to the same field shown in (a). Note that the contours of panel (a) are overplotted in (b) and (c) for clarity. (d) Optical \textit{HST} image of HCG 31 - a zoom from the larger region shown in (e). The green boxes on (d) correspond to the vertical panels on the left, labeled 1 to 7, which indicate the regions for which we have extracted the H$\alpha$ profiles (from the Fabry-Perot data). These profiles are normalized to the size of the box. Note that in black we show the integrated profile and in colors we highlight the components for which we can follow the velocity shift (left to right), from top to bottom (North to South).}
\label{profiles}
\vspace{1.cm}
\end{figure*}

\subsection{Kinematics: Using high-resolution Fabry-Perot data to search for gas flows}

Fig. \ref{profiles}a shows the total flux (in arbitrary units) within the interference filter of the Fabry-Perot, around Halpha (20 Angstrons), roughly matching the HST image displayed in (d). The bar-like structure of HCG 31A (eastern part of the galaxy) can be readily seen here.

Fig. \ref{profiles}b shows the velocity field of the same region, adapted from Amram et al. (2007). The velocity gradient in the region of interest (which corresponds to the location of panels 1 to 7 in panel d) is essentially flat. In particular, the two main clumps of star formation (HCG 31C and A) display quite similar radial velocities. Indeed, the barycenter of the entire profile is used to compute the mean velocity displayed in the velocity field, and this quantity displays a small global shift. This is due to the fact that, along the axis connecting the two coalescing regions, the relative intensity variation of each individual velocity component, pointed out in panel 1 to 7, roughly compensates the velocity shift during the computation of the barycentric radial velocity. The velocity gradient is thus smooth, without evidences of a kinematical feature due to different velocities in the two main blobs. Nevertheless, as will be described below, there is a clear shift in velocity seen for each individual velocity component that forms the H$\alpha$ profile and this can not be seen in the global radial velocity pattern of the system.

Fig. \ref{profiles}c shows the velocity dispersion map, adapted from Fig. 8 of Amram et al. (2007). The dispersion is very high in the merger-interface region (red and pink regions in the map). The contrast between the high- and low-dispersion regions is even more obvious when looking at the large scale figure given in Fig. 8 of Amram et al. (2007), which also shows that some other high-dispersion regions exist. The high values in the velocity dispersion map can be explained by the broadening of the H$\alpha$ profiles due to the presence of the several components as those displayed on panels 1-7. Outside this region the velocity dispersion values are lower (from blue to yellow, in the map). We do not see any spatial coincidence between the high-velocity dispersion region in panel 4c and the regions of highest intensity in panel 4a,  suggesting that the broadening of the profiles is not only linked with star formation, otherwise we would expect a direct correlation. We then speculate that there is room for the presence of a flow that carries the gas.

We have used the package {\sc pan}\footnote{http://ifs.wikidot.com/pan} in {\sc idl} to fit multiple Gaussians to the observed Fabry-Perot H$\alpha$ profiles of the central regions of HCG 31AC. Given the high spectral resolution of the Fabry-Perot observations and the complexity of the system, it is not possible to have a unique solution for the multiple Gaussian fitting process. In the fit we performed, we attempted to use the smallest number of Gaussians, and this resulted in five main components, which came out naturally from the best-fit profile decomposition. Center, width and intensity were free parameters.

In panels (d) and (e) of Figure \ref{profiles} we show an archival ACS \textit{Hubble Space Telescope} (\textit{HST}) image of HCG 31 (filter F435W). In (d) the green squares (0.8''$\times$0.8'' or 208$\times$208 pc each) represent the regions for which we have extracted H$\alpha$ profiles (these are shown as black continuum lines in panels 1 to 7). With color Gaussians of different intensities we show the several velocity components. We claim that these are moving from left to right as we inspect the panels from top to bottom (North to South). In the following, we exemplify our claim using two components, green and red. Both components can be readly seen in panel 1, with central velocities 3957 km s$^{-1}$ and 3978 km s$^{-1}$ respectively. In panel 2, we see the green component decrease its intensity by about 50\% and the red component increase its intensity by $\sim$10 - 20$\%$ and they shift by 4 km s$^{-1}$ and 15 km s$^{-1}$ to the right, respectively. In panel 3 the red component remains at the same radial velocity and the green component is shifted by 6 km s$^{-1}$ to the right, but with lower intensity. The scenario is different for panel 4. Inspecting the optical \textit{HST} image at this location, we can see a region of lower flux in between the two clumps.
Nevertheless the H$\alpha$ profile is high - it has similar intensities to panels 1 and 2. In panel 4 the green and red components are shifted to the right by $\sim$15 km s$^{-1}$ and $\sim$7 km s$^{-1}$ respectively, and the intensity of both components decrease, as compared to panel 3. In panel 5, the intensity of the green component increases but its velocity stays unchanged. The red component has its intensity increased by about a factor two and moves by $\sim$11 km s$^{-1}$ to the right. The green component is again not shifted in panel 6 and the red component moves $\sim$2 km s$^{-1}$. In panel 7 the green and red components move to the right by $\sim$10 km s$^{-1}$ and $\sim$4 km s$^{-1}$ respectively. Thus, from panel 1 to 7 the green and red components move a total of $\sim$33 km s$^{-1}$ and $\sim$37 km s$^{-1}$ respectively. All the above descriptions suggest that an ionized gas component is moving along the ridge that traces the merger interface between A and C. In general, all components show a regular shift to higher velocities, from North to South.

In conclusion, the gas flow we detect is widespread, traced by the high-velocity dispersion regions of the map, and in particular it is present in the merger interface region, along the line that joins the two coalescing nuclei.

\section{Discussion and summary}
\label{summary}

In this Letter we report, for the first time, direct observational evidence of metal mixing in merging galaxies. We detect a smooth metallicity distribution between two strong bursts of star formation, which are associated with two different interacting galaxies, namely, HCG 31A and HCG 31C. The analysis of high-resolution H$\alpha$ Fabry-Perot data reveals the presence of multiple H$\alpha$ emission line components in the merger interface, along the line that joins HCG 31A to HCG 31C main star forming complexes. A systematic shift in the radial velocity of these components supports the idea of gas flows. This gas flow should be the main responsible in producing a smooth metallicity connection between these objects. This observational result is in agreement with recent results obtained in simulations, i.e., interactions and mergers induce gas inflows towards the center of the galaxies, diluting the central abundances and producing flat metallicity gradients (Rupke et al 2010, Perez et al. 2011). In addition, strong bursts of star formation are triggered, as found in recent simulations (Hopkins et al. 2013). We are witnessing this process in an early-stage merger, where the main burst associated with each galaxy still retains its original metallicity, but a smooth gradient and fairly flat connects both systems. 

\acknowledgments

We warmly thank Thierry Contini, Marianne Girard, Enrique P\'erez-Montero, Carolina Kehrig, Angel L\'opez-S\'anchez and Raffaela Marino for insightful discussions that  improved this manuscript and Enrique P\'erez-Montero for making his code {\sc hi-chi-mistry} available for the community. Based on observations made with the NASA/ESA Hubble Space Telescope, obtained from the data archive at the Space Telescope Science Institute. STScI is operated by the Association of Universities for Research in Astronomy, Inc. under NASA contract NAS 5-26555. ST-F acknowledges the financial support of FONDECYT through a project ``Iniciaci\'on en la Investigaci\'on'', under contract 11121505 and the support of the project CONICYT PAI/ACADEMIA 7912010004. CMdO \& PA thank the support of USP/COFECUB. CMdO acknowledges support from FAPESP and CNPq. MAC acknowledges the financial support of the Direcci\'on de Investigaci\'on of the Universidad de La Serena, through a ``Concurso de Apoyo a Tesis 2013'', number PT13146.



{\it Facilities:} \facility{Gemini}, \facility{ESO}.

\clearpage

\end{document}